\documentclass{article}
\usepackage{spconf,amsmath,graphicx,amssymb}
\usepackage{algorithm,algorithmicx,algpseudocode}
\usepackage{amsfonts}
\usepackage{bm}
\usepackage{multirow}
\usepackage{booktabs}
\usepackage{breqn}
\usepackage{url}
\usepackage{arydshln}
\usepackage{setspace}
\usepackage{cite}

\newcommand{\dashrule}[1][black]{%
  \color{#1}\rule[\dimexpr.5ex-.2pt]{4pt}{.4pt}\xleaders\hbox{\rule{4pt}{0pt}\rule[\dimexpr.5ex-.2pt]{4pt}{.4pt}}\hfill\kern0pt%
}
\DeclareMathOperator*{\argmax}{argmax}

\makeatletter
\def\adl@drawiv#1#2#3{%
        \hskip.5\tabcolsep
        \xleaders#3{#2.5\@tempdimb #1{1}#2.5\@tempdimb}%
                #2\z@ plus1fil minus1fil\relax
        \hskip.5\tabcolsep}
\newcommand{\cdashlinelr}[1]{%
  \noalign{\vskip\aboverulesep
           \global\let\@dashdrawstore\adl@draw
           \global\let\adl@draw\adl@drawiv}
  \cdashline{#1}
  \noalign{\global\let\adl@draw\@dashdrawstore
           \vskip\belowrulesep}}
\makeatother

\title{Improved Mask-CTC for Non-Autoregressive End-to-End ASR}
\name{
    Yosuke Higuchi$^{1}$,
    Hirofumi Inaguma$^{2}$,
    Shinji Watanabe$^{3}$, 
    Tetsuji Ogawa$^{1}$,
    Tetsunori Kobayashi$^{1}$
}

\address{
    $^{1}$ Waseda University, Japan
    $^{2}$ Kyoto University, Japan
    $^{3}$ Johns Hopkins University, USA
 }

\begin{document}
\maketitle
\ninept
\setlength{\abovedisplayskip}{2mm}
\setlength{\belowdisplayskip}{2mm}

\begin{abstract}
For real-world deployment of automatic speech recognition (ASR), 
the system is desired to be capable of fast inference while relieving the requirement of computational resources. 
The recently proposed end-to-end ASR system based on mask-predict with connectionist temporal classification (CTC), \textit{Mask-CTC}, fulfills this demand by generating tokens in a non-autoregressive fashion. 
While Mask-CTC achieves remarkably fast inference speed, 
its recognition performance falls behind that of conventional autoregressive (AR) systems.
To boost the performance of Mask-CTC, 
we first propose to enhance the encoder network architecture by employing a recently proposed architecture called Conformer.
Next, we propose new training and decoding methods by introducing auxiliary objective to predict the length of a partial target sequence, which allows the model to delete or insert tokens during inference.
Experimental results on different ASR tasks show that 
the proposed approaches improve Mask-CTC significantly, 
outperforming a standard CTC model (15.5\% $\rightarrow$ 9.1\% WER on WSJ). 
Moreover, Mask-CTC now achieves competitive results to AR models with no degradation of inference speed ($<$ 0.1 RTF using CPU). 
We also show a potential application of Mask-CTC to end-to-end speech translation.
% Our codes will be publicly available on ESPnet\footnote{\url{https://github.com/espnet/espnet}}.
\end{abstract}
% 200 words


\vspace{-1mm}
\section{Introduction}
\vspace{-1mm}
End-to-end automatic speech recognition (E2E-ASR) is a task which directly converts speech into text based on sequence-to-sequence modeling~\cite{graves2006connectionist,  graves2012sequence, sutskever2014sequence, bahdanau2014neural}.
Compared with the traditional hybrid system built upon separately trained modules~\cite{hinton2012deep}, 
E2E-ASR greatly simplifies its model training and inference, 
reducing the cost of model development and the requirement of computational resources. 
Besides, E2E-ASR has achieved comparable results with those of the hybrid system in diverse tasks~\cite{chiu2018state, luscher2019rwth, karita2019a, sainath2020streaming}. 
Recent model architectures and techniques have further enhanced the performance of E2E-ASR~\cite{dong2018speech, sainath2019two, karita2019improving, kriman2020quartznet, han2020contextnet, gulati2020conformer}.

Many of the previous studies on E2E-ASR focus on an \textit{autoregressive} (AR) model~\cite{sutskever2014sequence, bahdanau2014neural}, 
which estimates the likelihood of a sequence generation
based on a left-to-right probabilistic chain rule. 
The AR model often performs the best among other E2E-ASR architectures~\cite{chiu2018state}. 
However, it suffers from slow inference speed, requiring $L$-step incremental calculations of the model to generate $L$ tokens.
While some neural models, including Transformer~\cite{vaswani2017attention}, permit for efficient inference using GPU, 
it becomes highly computation intensive in production environments. 
For example, mobile devices are often constrained by limited on-device resources, where only CPU is generally available for the model inference\footnote{It may be argued that the inference can be done on a cloud server but running the model entirely on-device is important from privacy perspective~\cite{sainath2020streaming}.}.
If the inference algorithm is light and simple, 
ASR can be performed on-device at low-power consumption with fast recognition speed.
Building E2E-ASR models with fast inference speed while suppressing the model calculations is a critical factor for the real-world deployment.

Contrary to the AR framework, 
a \textit{non-autoregressive} (NAR) model generates a sequence within a constant number of the inference steps~\cite{graves2006connectionist, gu2017non}.
%reducing the computational costs and thus speeding up the inference. 
NAR models in neural machine translation have achieved competitive performance to AR models~\cite{libovicky2018end, lee2018deterministic, stern2019insertion, gu2019levenshtein, ghazvininejad2019mask, ghazvininejad2020semi, saharia2020non, ma2019flowseq}.
% such as approaches based on 
% the iterative refinement decoding~\cite{lee2018deterministic}, 
% insert or edit-based sequence generation~\cite{stern2019insertion, gu2019levenshtein}, 
% masked language model (MLM) objective~\cite{ghazvininejad2019mask, ghazvininejad2020semi, saharia2020non}, and
% generative flow~\cite{ma2019flowseq}.
Several approaches have been proposed to realize a NAR model in E2E-ASR~\cite{graves2014towards, chen2019non, chan2020imputer, higuchi2020mask, fujita2020insertion, tian2020spike}.
Connectionist temporal classification (CTC) predicts a frame-wise latent alignment between input speech frames and output tokens, and
generates a target sequence based on a conditional independence assumption between the frame predictions~\cite{graves2014towards}.
However, such assumption limits the performance of CTC compared with that of AR models~\cite{battenberg2017exploring}. 
Imputer~\cite{chan2020imputer} effectively models the contextual dependencies by iteratively predicting the latent alignments of CTC based on mask prediction~\cite{devlin2019bert, ghazvininejad2019mask}. 
%During inference, a target sequence is generated in a constant number of steps by iteratively predicting partial alignments via mask prediction.
Despite achieving comparable performance with that of AR models, 
Imputer processes the \textit{frame-level} sequence, consisting of hundreds of units, using the self-attention layers~\cite{vaswani2017attention}, 
which cost computations proportional to the square of a sequence length.
On the other hand, Mask-CTC~\cite{higuchi2020mask} generates a sequence shorter than Imputer  
by refining a \textit{token-level} CTC output with the mask prediction, 
achieving fast inference speed of less than 0.1 real time factor (RTF) using CPU.

In this work, 
we focus on improving Mask-CTC based E2E-ASR, 
which suits for the production owing to its fast inference. 
Mask-CTC starts a sequence generation from an output of CTC to avoid the cumbersome length prediction of a target sequence, which is required in the previous NAR framework~\cite{gu2017non}. 
However, Mask-CTC is confronted with some problems related to this dependence on the CTC output. 
Firstly, 
the performance of Mask-CTC is strongly limited to that of CTC 
since only minor errors in the CTC output, such as spelling mistakes, are subjected to the refinement.
Secondly, the length of a target sequence is fixed with that of the CTC output throughout the decoding, 
making it difficult to recover deletion or insertion errors in the CTC output.
To overcome the former, 
we adopt Conformer~\cite{gulati2020conformer}, which yields better speech processing than Transformer, 
to improve the CTC performance.
For the latter, we introduce new training and decoding strategies to handle the deletion and insertion errors during inference.
We also explore the potential of Mask-CTC to the end-to-end speech translation (E2E-ST) task.

\vspace{-3mm}
\section{Non-autoregressive End-to-End ASR}
\vspace{-2mm}
E2E-ASR aims to model a conditional probability $P(Y|X)$, 
which directly maps
a $T$-length input sequence $X = (\bm{\mathrm{x}}_t \in \mathbb{R}^D| t=1,...,T)$ into 
a $L$-length output sequence $Y = ( y_l \in \mathcal{V} | l=1,...,L )$.
Here, 
$\bm{\mathrm{x}}_t$ is a $D$-dimensional acoustic feature at frame $t$, 
$y_l$ is an output token at position $l$, and 
$\mathcal{V}$ is a vocabulary.

In this section, 
we review \textit{non-autoregressive} (NAR) models, which perform sequence generation within a constant number of inference steps in parallel with token positions.
% independent on the length of a target sequence.

% \subsection{Connectionist temporal classification (CTC)}
% CTC predicts a frame-level alignment between the input $X$ and the output $Y$ by introducing a special blank symbol $\epsilon$.
% The alignment $A=\{ a_t \in \mathcal{V} \cup \{\epsilon\} | t=1,...,T\}$ is predicted from the input $X$ with the conditional independence assumption between the alignment predictions as follows:
% \begin{equation}
%     P_{\mathrm{align}} (A | X) = \prod_{t=1}^{T} P (a_t | X).
% \end{equation}
% CTC models the joint probability $P(Y|X)$ by marginalizing $Y$ over the alignment $A$ as:
% \begin{equation}
%     \label{eq:p_ctc}
%     P_{\mathrm{ctc}} (Y | X) = \sum_{A \in \beta^{-1} (Y)} P_{\mathrm{align}} (A | X),
% \end{equation}
% where $\beta^{-1} (Y)$ denotes all possible alignments compatible with $Y$.
% The conditional independence assumption permits the CTC objective to be trained efficiently via dynamic programming and 
% enables the model to perform non-autoregressive generation of a sequence in one decoding step.
\vspace{-2mm}
\subsection{Connectionist temporal classification (CTC)}
\vspace{-1mm}
CTC predicts a frame-level input-output alignment $A=(a_t \in \mathcal{V} \cup \{\epsilon\} | t=1,...,T)$ by introducing a special blank symbol $\epsilon$~\cite{graves2006connectionist}.
% between the input $X$ and the output $Y$
Based on a conditional independence assumption per frame,
% in the alignment prediction $P(A|X)$
CTC models the probability $P(Y|X)$ by marginalizing over all possible paths as:
\begin{equation}
    \label{eq:p_ctc}
    P_{\mathrm{ctc}} (Y | X) = \sum_{A \in \beta^{-1} (Y)} P (A | X),
\end{equation}
where $\beta^{-1} (Y)$ denotes all possible alignments compatible with $Y$.
The conditional independence assumption enables the model to perform NAR sequence generation in a single forward pass. %in one inference step.
% permits the CTC objective to be trained efficiently via dynamic programming and 

\vspace{-2mm}
\subsection{Mask-CTC} \label{ssec:maskctc}
\vspace{-1mm}
Due to the conditional independence assumption, 
the CTC-based model generally suffers from poor recognition performance~\cite{chiu2018state}.
Mask-CTC has been proposed to mitigate this problem by iteratively refining an output of CTC with bi-directional contexts of tokens~\cite{higuchi2020mask}. 
%
% attention-based 
Mask-CTC adopts an encoder-decoder model built upon Transformer blocks~\cite{vaswani2017attention}. CTC is applied to the encoder output~\cite{kim2017joint, karita2019improving} and the decoder is trained via the masked language model (MLM) objective~\cite{devlin2019bert, ghazvininejad2019mask}.
% In the training of the MLM decoder, 
For training the MLM decoder, 
randomly sampled tokens $Y_{\mathrm{mask}}$ are replaced by a special mask token \texttt{<MASK>}.
% simultaneously 
$Y_{\mathrm{mask}}$ are then predicted conditioning on the rest observed (unmasked) tokens $Y_{\mathrm{obs}}$ $(= Y \setminus Y_{\mathrm{mask}})$ as follows:
% the encoder output and 
\begin{equation}
      P_{\mathrm{mlm}} (Y_{\mathrm{mask}} | Y_{\mathrm{obs}}, X) = 
      \prod_{y \in Y_{\mathrm{mask}}} P (y | Y_{\mathrm{obs}}, X).
      \label{eq:p_mlm}
\end{equation}
The number of masked tokens $N_{\mathrm{mask}}$ $(= |Y_{\mathrm{mask}}|)$ are sampled from a uniform distribution of $1$ to $L$ as in~\cite{ghazvininejad2019mask, ghazvininejad2020semi}.
With the CTC objective in Eq. (\ref{eq:p_ctc}), Mask-CTC optimizes model parameters by maximizing the following log-likelihood as:
\begin{multline}
    \label{eq:l_nar}
    \mathcal{L}_{\mathrm{NAR}} = \alpha \log P_{\mathrm{ctc}} (Y | X) + \\
    (1 - \alpha) \log P_{\mathrm{mlm}} (Y_{\mathrm{mask}} | Y_{\mathrm{obs}}, X),
\end{multline}
where $\alpha$ $(0 \leq \alpha \leq 1)$ is a tunable parameter.

During inference, 
an output of CTC is obtained through greedy decoding by suppressing repeated tokens and removing blank symbols.
Then, low-confidence tokens are replaced with \texttt{<MASK>} 
by thresholding the posterior probabilities of CTC with $P_{\mathrm{thres}}$.
The resulting unmasked tokens are then fed into the MLM decoder to predict the masked tokens. 
With this two-pass inference based on the CTC and MLM decoding, 
the errors in the CTC output, caused by the conditional independence assumption, are expected to be corrected conditioning on the high-confidence tokens in the entire output sequence.
The CTC output can be further improved by gradually filling in the masked tokens in multiple steps $K$.
At the $n$-th step, 
$y_{l} \in Y_{\mathrm{mask}}^{(n)}$ is predicted as follows:
\begin{equation}
    \label{eq:predict}
    y_l= \argmax_{y} P_{\mathrm{mlm}} (y_l = y | Y_{\mathrm{obs}}^{(n)}, X),
\end{equation}
where top $C$ positions with the highest decoder probabilities are selected to be predicted at each iteration.
By setting $C = \lfloor N_{\mathrm{mask}} / K \rfloor$, 
inference can be completed in constant $K$ steps.

\section{Proposed Improvements of Mask-CTC}
\vspace{-1mm}
Making use of the CTC output during inference, 
Mask-CTC effectively avoids the cumbersome length prediction of a target sequence~\cite{gu2017non},
which is rather challenging in ASR~\cite{chen2019non, higuchi2020mask}.
However, 
there are some drawbacks to this dependence on the CTC output. 
Firstly, 
% strongly 
the performance of CTC limits that of Mask-CTC 
because the MLM decoder can only make minor changes to the CTC output, such as correcting spelling errors.
Secondly, as the target length is fixed with that of the CTC output throughout the decoding, 
it is difficult to recover deletion or insertion errors. 
% Accordingly, the recognition performance of Mask CTC falls behind that of the autoregressive model~\cite{higuchi2020mask}.

To tackle these problems, 
we propose to (1) enhance the encoder architecture by adopting the state-of-the-art encoder architecture, Conformer~\cite{gulati2020conformer}, and 
% to boost the CTC performance
(2) introduce new training and decoding methods for the MLM decoder to handle the insertion and deletion errors during inference.

% To tackle this problem, 
% we improve the MLM decoder by introducing a training task for predicting the ``length'' of each masked token and a decoding algorithm to generate a sequence with a flexible length.
% We also employ Conformer~\cite{gulati2020conformer}, 
% the state-of-the-art E2E-ASR encoder architecture, 
% to boost the performance of CTC and thus improve the overall performance of Mask CTC.
\vspace{-2.5mm}
\subsection{Conformer}
\vspace{-1mm}
Conformer integrates a convolution module and Macaron-Net style feed-forward network (FFN)~\cite{lu2019understanding} into the Transformer encoder block~\cite{vaswani2017attention}.
% and mainly comprises of a convolution and self-attention modules.
While the self-attention layer is effective at modeling long-range global context, 
the convolution layer increases the ability to capture local patterns in a sequence, which is effective for speech processing.
We expect the Conformer encoder to boost the performance of CTC and thus improve the overall performance of Mask-CTC accordingly.

\vspace{-2.5mm}
\subsection{Dynamic length prediction (DLP)}\label{ssec:dlp}

\subsubsection{Training}
\vspace{-1mm}
%In order to make the MLM decoder capable of recovering insertion and deletion errors during inference, 
Inspired by~\cite{ren2019fastspeech, gu2019levenshtein},  
we propose dynamic length prediction (DLP), in which the MLM decoder is trained so as to dynamically predict the length of a partial target sequence from a corresponding masked token at each iteration.
In addition to the mask prediction task in Eq. (\ref{eq:p_mlm}), which is analogous to solving substitution errors, 
the decoder is trained to solve simulated deletion and insertion errors.
% solve tasks simulated for detecting deletion and insertion errors.

\vspace{1mm} \noindent \textbf{Deletion-simulated task} makes the decoder predict the length of a partial target sequence, having ``one or more" tokens, from a corresponding masked token.
For example, 
given a ground-truth sequence $Y = [y_1, y_2, y_3, y_4]$ and 
its masked sequence $[y_1, \text{\texttt{<MASK>}}, y_4]$ by merging $y_2$ and $y_3$ to \texttt{<MASK>}, 
the decoder is trained to predict a symbol \texttt{2} from the masked position, which corresponds to the length of the partial sequence $[y_2, y_3]$.
% in $Y$
This task makes the decoder aware of the possibility that the masked token has one or more corresponding output tokens, 
simulating a recover from deletion error in the decoder inputs.
To generate such masks, 
partial tokens in $Y$ are randomly sampled and replaced with {\texttt{<MASK>}} as $Y_{\mathrm{mask}}$ in Eq. (\ref{eq:p_mlm}). 
Then the consecutive masks are integrated into one single mask to form a masked sequence $Y^{\mathrm{del}}$ ($|Y^{\mathrm{del}}| \le |Y|$), consisting of 
masked tokens $Y_{\mathrm{mask}}^{\mathrm{del}}$ and 
observed tokens $Y_{\mathrm{obs}}^{\mathrm{del}}$ $(=  Y^{\mathrm{del}} \setminus Y_{\mathrm{mask}}^{\mathrm{del}})$.
The target length labels, $D_{\mathrm{del}} = \{d_i \in \mathcal{Z} | i=1, ..., |Y_{\mathrm{mask}}^{\mathrm{del}}|\}$, are obtained from the above mask integration process.
$D_{\mathrm{del}}$ is predicted conditioning on the observed tokens $Y_{\mathrm{obs}}^{\mathrm{del}}$ as:
% and the input speech $X$ 
\begin{equation}
    P_{\mathrm{lp}}(D_{\mathrm{del}} | Y_{\mathrm{obs}}^{\mathrm{del}}, X) = \prod_{i} P (d_i | Y_{\mathrm{obs}}^{\mathrm{del}}, X),
\end{equation}
where $P_{\mathrm{lp}}$ is a posterior probability distribution of the mask length.

\vspace{1mm} \noindent \textbf{Insertion-simulated task} makes the decoder predict ``zero'' from a masked token, 
indicating the mask corresponds to no partial target sequence.
For example, 
given a ground-truth sequence $Y = [y_1, y_2, y_3, y_4]$ and 
a masked sequence $[y_1, y_2, y_3, \text{\texttt{<MASK>}}, y_4]$, %by inserting \texttt{<MASK>}, 
the decoder is trained to predict a symbol \texttt{0} from the masked position as there is no corresponding tokens in $Y$.
This way, we can make the decoder aware of the redundant masked token, 
simulating a recover from insertion error in the decoder inputs.
To obtain masks for this task, 
we randomly insert {\texttt{<MASK>}} into $Y$ to form a masked sequence $Y^{\mathrm{ins}}$ ($|Y^{\mathrm{ins}}| > |Y|$), 
consisting of 
masked tokens $Y_{\mathrm{mask}}^{\mathrm{ins}}$ and 
observed tokens $Y_{\mathrm{obs}}^{\mathrm{ins}}$ $(= Y^{\mathrm{ins}} \setminus Y_{\mathrm{mask}}^{\mathrm{ins}})$.
The target lengths $D_{\mathrm{ins}} = \{d_i = 0 | i = 1, ..., |Y_{\mathrm{mask}}^{\mathrm{ins}}|\}$ of the inserted masks are predicted as:
\begin{equation}
    P_{\mathrm{lp}}(D_{\mathrm{ins}} | Y_{\mathrm{obs}}^{\mathrm{ins}}, X) = \prod_{i} P (d_i | Y_{\mathrm{obs}}^{\mathrm{ins}}, X).
\end{equation}

Both tasks are trained jointly by a shared single linear layer followed by a softmax classifier (we set the maximum class to 50) on top of the decoder.
The objective of the overall DLP is formulated by combining Eqs. (5) and (6) as:
\begin{equation}
    \label{eq:l_lp}
    \mathcal{L}_{\mathrm{LP}} = 
    \log P_{\mathrm{lp}}(D_{\mathrm{del}} | Y_{\mathrm{obs}}^{\mathrm{del}}, X) + 
    \log P_{\mathrm{lp}}(D_{\mathrm{ins}} | Y_{\mathrm{obs}}^{\mathrm{ins}}, X).
\end{equation}
Finally, a new Mask-CTC model is trained with a loss combining the conventional objective $\mathcal{L}_{\mathrm{NAR}}$ from Eq. (\ref{eq:l_nar}) and the objective of the proposed DLP $\mathcal{L}_{\mathrm{LP}}$ from Eq. (\ref{eq:l_lp}) as follows:
\begin{equation}
    \label{eq:ctc-cmlm-lp}
    \mathcal{L}_{\mathrm{NAR-LP}} = 
    \mathcal{L}_{\mathrm{NAR}} + 
    \beta \mathcal{L}_{\mathrm{LP}},
\end{equation}
where $\beta$ $(> 0)$ is a tunable parameter.

\vspace{-2mm}
\subsubsection{Inference}\label{sssec:sedecoding}
\vspace{-1mm}
\begin{algorithm}[t]
    \setstretch{1.1}
    \footnotesize
    \caption{\bf Shrink-and-Expand Decoding}
    \begin{algorithmic}[1]
    \renewcommand{\algorithmicrequire}{\textbf{Input:}}
    \renewcommand{\algorithmicensure}{\textbf{Output:}}
        \Require $K$: iteration step, $X$: intput speech, $\hat{Y} = \{\hat{y}_l\}_{l=1}^{L}$: CTC output
        %\Ensure $Y$: generated sequence
        \State Calculate $P_{l, \mathrm{mlm}} = P_{\mathrm{mlm}} (y_l = \hat{y_l} | \hat{Y}, X)$ for each $\hat{y}_l \in \hat{Y}$
        \State Obtain $\hat{Y}_{\mathrm{mask}} = \{\hat{y}_l | P_{l, \mathrm{mlm}} < P_{\mathrm{thres}} \}_{l=1}^{L}$
        \State Mask $\hat{Y}$, where $\hat{Y} = 
        \begin{cases}
            \varnothing & (\hat{y}_l \in  \hat{Y}_{\mathrm{mask}}) \quad \text{\footnotesize \texttt{// <MASK>}}\\
            \hat{y}_l & (\hat{y}_l \in \hat{Y}_{\mathrm{obs}} = \hat{Y} \setminus \hat{Y}_{\mathrm{mask}}) 
        \end{cases}$
        \State $C = \lfloor |\hat{Y}_{\mathrm{mask}}| / K \rfloor$ {\footnotesize \texttt{// \#masks predicted in each loop}}
        \While {stopping criterion not met}
            \State {\scriptsize \texttt{// {\footnotesize Shrink}} {\footnotesize \texttt{(}}\texttt{Ex.} $[\hat{y}_1, \varnothing, \varnothing, \varnothing, \hat{y}_5, \varnothing, \hat{y}_7] \rightarrow [\hat{y}_1, \varnothing, \hat{y}_5, \varnothing, \hat{y}_7]$}{\footnotesize \texttt{)}}
            \State \textit{Shrink} masks in $\hat{Y}$ and update $\hat{Y}_{\mathrm{mask}}$, $\hat{Y}_{\mathrm{obs}}$ accordingly
            \Statex \vspace{-7pt}
            \State {\scriptsize \texttt{// {\footnotesize Expand}} {\footnotesize \texttt{(}}\texttt{Ex.} $[\hat{y}_1, \varnothing, \hat{y}_5, \varnothing, \hat{y}_7] \rightarrow [\hat{y}_1, \varnothing, \varnothing, \hat{y}_5, \hat{y}_7]$}{\footnotesize \texttt{)}}
            \State Calculate  $P_{l, \mathrm{lp}} (d_l | \hat{Y}_{\mathrm{obs}}, X)$ for each $\hat{y}_l \in  \hat{Y}_{\mathrm{mask}}$
            \State \textit{Expand} masks in $\hat{Y}$ based on $\argmax_{d} P_{l, \mathrm{lp}} (d_l =d | \hat{Y}_{\mathrm{obs}}, X)$
            \Statex \hspace{33pt}  and update $\hat{Y}_{\mathrm{mask}}$, $\hat{Y}_{\mathrm{obs}}$ accordingly
            \Statex \vspace{-7pt}
            \State Calculate $P_{l, \mathrm{mlm}} (y_l | \hat{Y}_{\mathrm{obs}}, X)$ for each $\hat{y}_l \in \hat{Y}_{\mathrm{mask}}$
            \State Predict masks in $\hat{Y}$
            as $\argmax_y P_{l, \mathrm{mlm}} (y_l = y| \hat{Y}_{\mathrm{obs}}, X)$, 
            \Statex \hspace{33pt} where $\hat{y}_l$ with top-$C$ highest probabilities are selected
            \State Update $\hat{Y}_{\mathrm{mask}}$, $\hat{Y}_{\mathrm{obs}}$
        \EndWhile \\
        \Return $\hat{Y}$
    \end{algorithmic}
\end{algorithm}
Alg. 1 shows the proposed \textit{shrink-and-expand decoding} algorithm, 
which allows delete and insert tokens during sequence generation.
Compared with the conventional method~\cite{higuchi2020mask} (explained in Sec. \ref{ssec:maskctc}), 
the proposed decoding differs in the masking process of the CTC output and 
the prediction process of the masked tokens.

While the previous method detects low-confidence tokens in a CTC output by thresholding the posterior probabilities of CTC, 
the proposed decoding refers to the probabilities of the MLM decoder (line 2 in Alg. 1).
Taking advantage of the bi-directional contexts of tokens, 
it appeared that the decoder probabilities are more suitable for detecting the CTC errors. 

Shrink-and-expand decoding dynamically changes the target sequence length by 
deleting or inserting masks at each iteration.
\textit{Shrink} step (line 7 in Alg. 1) integrates consecutive masks into one single mask.
The integrated masks are then fed to the decoder to predict the length of each mask (line 9 in Alg. 1) required to match the length of an expected target sequence.
\textit{Expand} step (line 10 in Alg. 1) modifies the number of each mask based on the predicted length.
For example, if the length of a mask is predicted as \texttt{2}, 
we insert a mask to form two consecutive masks,
and if predicted as \texttt{0}, 
we delete the mask from the target sequence.
Finally, the masked tokens are predicted as the previous way in Eq. (\ref{eq:predict}).
Note that shrink-and-expand decoding requires two forward calculations of the decoder for each inference step: one for predicting the length of each mask and one for predicting target tokens.

% Our approach can be similar to Levenshtein Transformer~\cite{gu2019levenshtein}, which uses a placeholder classifier and a deletion classifier to insert or delete tokens in a target sequence, and
% FastSpeech~\cite{ren2019fastspeech}, which uses a duration predictor to change the length of an input phoneme sequence.
% However, we exploit mask token to make those operations possible for the MLM-based model.

\vspace{-1mm}
\section{Experiments}
\vspace{-1.5mm}
\begin{table*}[t]
    \centering
    \caption{Word error rate (WER) and real time factor (RTF) on WSJ. The proposed improvements on Mask-CTC (the use of Conformer and dynamic length prediction) are compared with CTC and autoregressive (AR) models.
    $K$ denotes the number of inference steps required to generate each output token. RTF was measured on dev93 using CPU. For each Mask-CTC model, RTF was calculated for $K=10$.}
    \label{tb:wsj}
    \centering
    \vspace{1mm}
    % 0.85
    \scalebox{0.82}{
    \begin{tabular}{lccccccccccr}
    \toprule
    \multirow{2}{*}[-3pt]{\textbf{Model}} & \multirow{2}{*}[-3pt]{\shortstack{\textbf{\#Params}\\(M)}} & \multicolumn{8}{c}{\textbf{WSJ} (WER)} &  \multirow{2}{*}[-3pt]{\textbf{\qquad RTF \qquad}} & \multirow{2}{*}[-3pt]{\textbf{Speedup}} \\
    \cmidrule(lr{0.4em}){3-10}
    & & \multicolumn{4}{c}{dev93} & \multicolumn{4}{c}{eval92} & & \\
    \midrule
    \midrule
    % dev: 48734 / 503 = 99.887, eval: 33398 / 333 = 100.294
    \textbf{\textit{Autoregressive}} & & \multicolumn{4}{c}{$K \! = \! L$ (avg. 99.9)} & \multicolumn{4}{c}{$K \! = \! L$ (avg. 100.3)} & & \\ 
    \cmidrule{1-12}
    \texttt{A1\ } Transformer-AR~\cite{karita2019improving} & 27.2 & \multicolumn{4}{c}{13.5} & \multicolumn{4}{c}{10.8} & 0.456$_{\pm0.005}$ & 1.00$\times$\\
    \texttt{A2\ } \hspace{1mm} + beam search & 27.2 & \multicolumn{4}{c}{12.8} & \multicolumn{4}{c}{10.6} & 5.067$_{\pm0.012}$ & 0.09$\times$\\
    \texttt{A3\ } Conformer-AR & 30.4 & \multicolumn{4}{c}{11.4} & \multicolumn{4}{c}{8.8} & 0.474$_{\pm0.009}$ & 0.96$\times$\\
    \texttt{A4\ } \hspace{1mm} + beam search & 30.4 & \multicolumn{4}{c}{\textbf{11.1}} & \multicolumn{4}{c}{\textbf{8.5}} & 5.094$_{\pm0.031}$ & 0.09$\times$ \\
    \midrule
    \midrule
    \multirow{2}{*}[-3pt]{\textbf{\textit{Non-autoregressive}}} & & \multicolumn{4}{c}{$K \! \le C$ ($C$: const.)} & \multicolumn{4}{c}{$K \! \le C$ ($C$: const.)} & & \\
    \cmidrule(lr{0.4em}){3-6} \cmidrule(lr{0.4em}){7-10}
    & & 0 & 1 & 5 & 10 & 0 & 1 & 5 & 10 & \\
    \cmidrule{1-12}
    \texttt{B1\ } Transformer-CTC & 17.7 & 19.4 & -- & $-$ & $-$ & 15.5 & $-$ & $-$ & $-$ & 0.021$_{\pm0.000}$ & 21.71$\times$ \\
    \texttt{B2\ } Transformer-Mask-CTC~\cite{higuchi2020mask} & 27.2 & 15.5 & 15.2 & 14.9 & 14.9 & 12.5 & 12.2 & 12.0 & 12.0 & 0.063$_{\pm0.001}$ & 7.24$\times$ \\
    \texttt{B3\ } \hspace{1mm} + dynamic length prediction & 27.2 & 15.5 & 14.0 & 13.9 & \textbf{13.8} & 12.4 & 11.1 & \textbf{10.8} & \textbf{10.8} & 0.074$_{\pm0.001}$ & 6.16$\times$ \\
    \cdashlinelr{1-12}
    \texttt{C1\ } Conformer-CTC & 20.9 & 13.0 & $-$ & $-$ & $-$ & 10.8 & $-$ & $-$ & $-$ & 0.033$_{\pm0.000}$ & 13.81$\times$ \\
    \texttt{C2\ } Conformer-Mask-CTC & 30.4 & 11.9 & 11.8 & 11.7 & 11.7 & 9.4 & 9.2 & 9.2 & \textbf{9.1} & 0.063$_{\pm0.000}$ & 7.24$\times$ \\
    \texttt{C3\ } \hspace{1mm} + dynamic length prediction & 30.4 & 11.8 & \textbf{11.3} & \textbf{11.3} & \textbf{11.3} & 9.6 & 9.3 & \textbf{9.1} & \textbf{9.1} & 0.080$_{\pm0.000}$ & 5.70$\times$ \\
    \bottomrule
    \end{tabular}}
    \vspace{-7mm}
\end{table*}
\begin{table}[t]
    \centering
    %  test sets
    \caption{Word error rates (WER) on Voxforge Italian and TEDLIUM2. Results with beam search are reported in parentheses.}
    \label{tb:vox_ted}
    \vspace{1mm}
    % 0.85
    \scalebox{0.81}{
    \begin{tabular}{lcccc}
        \toprule
        \textbf{Model} & \textbf{\#itr} $K$ & \textbf{Voxforge} & \textbf{TEDLIUM2} \\
        \midrule
        TF-AR~\cite{karita2019improving} & $L$ & 35.5 (35.7) & 9.5 (8.9) \\
        CF-AR & $L$ & \textbf{29.8} (\textbf{29.8}) & 8.4 (\textbf{7.9}) \\
        \cmidrule{1-4}
        TF-CTC & 0 & 56.1 & 16.6 \\
        %TF-Mask CTC & 0 & 39.8 & 11.4 \\
        TF-Mask-CTC~\cite{higuchi2020mask} & 10 & 38.3 & 10.9 \\
        \hspace{1mm} + DLP & 5 & \textbf{35.1} & \textbf{10.6} \\
        \cdashlinelr{1-4}
        CF-CTC & 0 & 31.8 & 9.5 \\
        %CF-Mask CTC & 0 & 29.4 & 8.7 \\
        CF-Mask-CTC & 10 & 29.2 & \textbf{8.6} \\
        \hspace{1mm} + DLP & 5 & \textbf{29.0} & 9.7 \\
        \bottomrule
    \end{tabular}}
\end{table}

To evaluate the effectiveness of the proposed improvements of Mask-CTC, we conducted 
experiments on E2E-ASR models using ESPnet~\cite{watanabe2018espnet}.
The recognition performance and the inference speed were evaluated based on word error rate (WER) and real time factor (RTF), respectively.
All of the decodings were done without using external language models (LMs).

\vspace{-2.5mm}
\subsection{Datasets}
\vspace{-1mm}
The experiments were carried out using three tasks: the 81 hours Wall Street Journal (WSJ)~\cite{paul1992design}, the 210 hours TEDLIUM2~\cite{rousseau2014enhancing} and the 16 hours Voxforge in Italian~\cite{voxforge}.
For the network inputs, we used 80 mel-scale filterbank coefficients with three-dimensional pitch features extracted using Kaldi~\cite{povey2011kaldi}. 
To avoid overfitting, 
we chose data augmentation techniques from speed perturbation~\cite{ko2015audio} and SpecAugment~\cite{park2019specaugment}, depending on the tasks and models.
For the tokenization of target texts, we used characters for WSJ and Voxforge.
For TEDLIUM2, we generated a 500 subword vocabulary based on Byte Pair Encoding (BPE) algorithm~\cite{sennrich2016neural}.

\vspace{-2.5mm}
\subsection{Experimental setup}
\vspace{-1mm}
For all the tasks, 
we adopted the same Transformer~\cite{vaswani2017attention} architecture as in~\cite{karita2019improving}, 
where the number of heads $H$, the dimension of a self-attention layer $d^{\mathrm{att}}$, the dimension of a feed-forward layer $d^{\mathrm{ff}}$ were set to 4, 256, and 2048, respectively.
The encoder consisted of 2 CNN-based downsampling layers followed by 12 self-attention layers and 
the decoder consisted of 6 self-attention layers. 
For Conformer encoder~\cite{gulati2020conformer}, 
we used the same configuration for the self-attention layer as in Transformer, except $d^{\mathrm{ff}}$ was set to $1024$.\footnote{We adjusted $d_{\mathrm{ff}}$ to avoid the Conformer-based model becoming slow during inference due to the increase in number of parameters.}
%The kernel size was tuned from values of 7, 15, and 31, depending on the task. 
For the model training, 
we tuned hyper-parameters (e.g., training epochs, etc.) following the recipes provided by ESPnet and 
we will make our configurations publicly available on ESPnet to ensure reproducibility.
% tuned training epochs, a learning rate, and a mini-batch size based on the recipes provided by ESPnet for each task. 
The loss weight $\alpha$ in Eq. (\ref{eq:l_nar}) was set to 0.3 and $\beta$ in Eq. (\ref{eq:ctc-cmlm-lp}) was set to 0.1 for Voxforge and 1.0 for WSJ and TEDLIUM2.
After the training, a final model was obtained by averaging model parameters over 
the 10 -- 50 checkpoints with the best validation performance.
During inference, the threshold $P_{\mathrm{thres}}$ for the conventional Mask-CTC (explained in Sec. \ref{ssec:maskctc}) was tuned from values of 0.9, 0.99, 0.999. 
For the proposed decoding in Sec. \ref{ssec:dlp}, we fixed $P_{\mathrm{thres}}$ to 0.5 for all tasks. 
RTF was measured using utterances in dev93 of WSJ using Intel(R) Xeon(R) Gold 6148 CPU, 2.40GHz.

\vspace{-2mm}
\subsection{Evaluated models}
\vspace{-1mm}
We evaluated different E2E-ASR models, each of which can either be \textit{autoregressive} (AR) or \textit{non-autoregressive} (NAR).
\textbf{AR} indicates an AR model trained with the joint CTC-attention objective~\cite{kim2017joint, karita2019improving} and, for inference, we used the joint CTC-attention decoding~\cite{hori2017joint} with beam search.
\textbf{CTC} indicates a NAR model simply trained with the CTC objective as in Eq. (\ref{eq:p_ctc})~\cite{graves2014towards}. 
\textbf{Mask-CTC} indicates a NAR model trained with the conventional Mask-CTC framework~\cite{higuchi2020mask} as explained in Sec. \ref{ssec:maskctc}. We also applied the proposed dynamic length prediction (DLP, explained in Sec. \ref{ssec:dlp}) to \textbf{Mask-CTC}.

For each model, we compared the results between Transformer (\textbf{TF}) and Conformer (\textbf{CF}) encoders.

\vspace{-2mm}
\subsection{Results}
\vspace{-1mm}
Table \ref{tb:wsj} shows the results on WSJ. 
By comparing the results among NAR models using Transformer (\texttt{B*}), 
the proposed DLP (\texttt{B3}) outperformed the conventional Mask-CTC (\texttt{B2}). 
We can conclude that DLP successfully recovered insertion and deletion errors in the CTC output, looking at \texttt{B2} and \texttt{B3} having the same CTC results ($K=0$) and \texttt{B3} resulted in better improvement.
The performance of CTC significantly improved by using Conformer (\texttt{B1}, \texttt{C1}) and 
Mask-CTC greatly benefited from it (\texttt{C2}). 
The errors were further reduced by applying DLP (\texttt{C3}), 
achieving 9.1\% in eval92 which was the best among the NAR models and better than that of the state-of-the-art model without LM~\cite{Sabour2019OptimalCD, borgholt2020do}.
By comparing results between NAR and AR models, 
Mask-CTC achieved highly competitive performance to AR models for both Transformer (\texttt{A1}, \texttt{B3}) and Conformer (\texttt{A3}, \texttt{C3}), 
demonstrating the effectiveness of the proposed methods for improving the original Mask-CTC.

In terms of the decoding speed using CPU, 
all NAR models (\texttt{B*}, \texttt{C*}) were at least 5.7 times faster than the AR models (\texttt{A*}), 
achieving RTF of under 0.1. 
The speed of Mask-CTC slowed down by applying DLP (\texttt{B3}, \texttt{C3}) due to the increase in number of decoder calculations (explained in Sec. \ref{sssec:sedecoding}). 
However, with DLP, the error rates converged faster in less iterations ($K=5$) and thus the inference speed was the same or even faster than the original Mask-CTC.

Table \ref{tb:vox_ted} shows the results on Voxforge and TEDLIUM2. 
Mask-CTC achieved comparable results with those of AR models by the proposed improvements.
However, DLP did not improve CF-Mask-CTC on TEDLIUM2. 
We observed the performance of CF-Mask-CTC was accurate enough 
and DLP was not effective.
% worsened the training of decoder.

\vspace{-2mm}
\subsection{End-to-end speech translation (E2E-ST)}
\vspace{-1mm}
To see a potential application to other speech tasks, 
we applied Mask-CTC framework to the E2E-ST task, following \cite{inaguma2020espnet}.
For NAR models, we used sequence-level knowledge distillation~\cite{kim2016sequence}.
Table \ref{tab:result_speech_translation} shows the results on Fisher-CallHome Spanish corpus~\cite{fisher_callhome}.
% Here, the target for the CTC layer is a translation, not transcription.
Since input-output alignments are non-monotonic in this task, we observed the confidence filtering based on the CTC probabilities did not work well, unlike ASR.
Next, we performed the mask-predict (MP) decoding proposed in MT~\cite{ghazvininejad2019mask} by starting from all \texttt{<MASK>} and confirmed some gains over CTC.
Finally, we initialized a target sequence with the filtered CTC output as in the ASR task and then performed the MP decoding.
Here, the number of masked tokens at each iteration are truncated by the number of initial masked tokens $N_{\mathrm{mask}}$ (\textit{restricted MP}) to keep information from the CTC output for the later iterations.
This way, the results were further improved from the CTC greedy results by a large margin. 
Moreover, interestingly, Mask-CTC outperformed the AR model on this corpus.

\begin{table}[t]
    \centering
    \caption{Speech translation results on Fisher-CallHome Spanish}\label{tab:result_speech_translation}
    \vspace{1mm}
    % 0.85
    \scalebox{0.83}{
    \begin{tabular}{lccccc}\toprule
    \multirow{2}{*}[-5pt]{\textbf{Model}} & \multicolumn{3}{c}{\textbf{Fisher} (BLEU)} & \multicolumn{2}{c}{\textbf{CallHome} (BLEU)} \\  \cmidrule(lr){2-4} \cmidrule(lr){5-6}
       & dev & dev2 & test & \ devtest \ & evltest \\ \midrule
      TF-AR & 47.01 & 47.89 & 47.19 & 18.11 & 17.95 \\
      TF-CTC & 45.57 & 46.97 & 45.97 & 15.99 & 15.91 \\
      TF-Mask-CTC &  &  &  &  &  \\
      \ + CTC greedy & 45.93 & 46.82 & 46.17 & 15.73 & 15.60 \\
      \ + original decoding & 44.80 & 45.40 & 44.39 & 14.14 & 14.14 \\
      \ + mask-predict (MP) & 47.43 & 48.14 & 46.96 & 16.52 & 16.42 \\
      \ + restricted MP & \bf{49.94} & \bf{49.42} & \bf{48.66} & \bf{16.96} & \bf{16.79} \\
      \bottomrule
    \end{tabular}}
\end{table}

\vspace{-2mm}
\section{Conclusion}
\vspace{-1mm}
This paper proposed improvements of Mask-CTC based non-autoregressive E2E-ASR. 
We adopted the Conformer encoder to boost the performance of CTC and 
introduced new training tasks 
% to predict the length of masked tokens, 
% which allowed the model to dynamically delete or insert tokens during inference. 
for the model to dynamically delete or insert tokens during inference. 
The experimental results demonstrated the effectiveness of the improved Mask-CTC, achieving competitive performance to autoregressive models with no degradation of inference speed. 
We also showed Mask-CTC framework can be used for end-to-end speech translation. 
Our future plan is to integrate an extrenal language model into Mask-CTC while keeping the decoding speed fast.

\newpage
%\small
\begin{spacing}{0.78}
\bibliographystyle{IEEEbib}
\setlength\itemsep{-1mm}
\bibliography{refs}

\begin{thebibliography}{10}

\bibitem{graves2006connectionist}
Alex Graves et~al.,
\newblock ``Connectionist temporal classification: labelling unsegmented
  sequence data with recurrent neural networks,''
\newblock in {\em Proc. of ICML}, 2006, pp. 369--376.

\bibitem{graves2012sequence}
Alex Graves,
\newblock ``Sequence transduction with recurrent neural networks,''
\newblock {\em arXiv preprint arXiv:1211.3711}, 2012.

\bibitem{sutskever2014sequence}
Ilya Sutskever and other,
\newblock ``Sequence to sequence learning with neural networks,''
\newblock in {\em Proc. of NeurIPS}, 2014, pp. 3104--3112.

\bibitem{bahdanau2014neural}
Dzmitry Bahdanau et~al.,
\newblock ``Neural machine translation by jointly learning to align and
  translate,''
\newblock in {\em Proc. of ICLR}, 2015.

\bibitem{hinton2012deep}
Geoffrey Hinton et~al.,
\newblock ``Deep neural networks for acoustic modeling in speech recognition:
  The shared views of four research groups,''
\newblock {\em IEEE Signal processing magazine}, vol. 29, no. 6, pp. 82--97,
  2012.

\bibitem{chiu2018state}
Chung-Cheng Chiu et~al.,
\newblock ``State-of-the-art speech recognition with sequence-to-sequence
  models,''
\newblock in {\em Proc. of ICASSP}, 2018, pp. 4774--4778.

\bibitem{luscher2019rwth}
Christoph L{\"u}scher et~al.,
\newblock ``{RWTH} {ASR} systems for librispeech: Hybrid vs attention,''
\newblock in {\em Proc. of Interspeech}, 2019, pp. 231--235.

\bibitem{karita2019a}
Shigeki Karita et~al.,
\newblock ``A comparative study on {Transformer} vs {RNN} in speech
  applications,''
\newblock in {\em Proc. of ASRU}, 2019, pp. 449--456.

\bibitem{sainath2020streaming}
Tara~N Sainath et~al.,
\newblock ``A streaming on-device end-to-end model surpassing server-side
  conventional model quality and latency,''
\newblock in {\em Proc. of ICASSP}, 2020, pp. 6059--6063.

\bibitem{dong2018speech}
Linhao Dong et~al.,
\newblock ``Speech-{Transformer}: a no-recurrence sequence-to-sequence model
  for speech recognition,''
\newblock in {\em Proc. of ICASSP}, 2018, pp. 5884--5888.

\bibitem{sainath2019two}
Tara~N Sainath et~al.,
\newblock ``Two-pass end-to-end speech recognition,''
\newblock in {\em Proc. of Interspeech}, 2019, pp. 2773--2777.

\bibitem{karita2019improving}
Shigeki Karita et~al.,
\newblock ``Improving {Transformer}-based end-to-end speech recognition with
  connectionist temporal classification and language model integration,''
\newblock in {\em Proc. of Interspeech}, 2019, pp. 1408--1412.

\bibitem{kriman2020quartznet}
Samuel Kriman et~al.,
\newblock ``Quartznet: Deep automatic speech recognition with 1d time-channel
  separable convolutions,''
\newblock in {\em Proc. of ICASSP}, 2020, pp. 6124--6128.

\bibitem{han2020contextnet}
Wei Han et~al.,
\newblock ``{ContextNet}: Improving convolutional neural networks for automatic
  speech recognition with global context,''
\newblock in {\em Proc. of Interspeech}, 2020.

\bibitem{gulati2020conformer}
Anmol Gulati et~al.,
\newblock ``Conformer: Convolution-augmented {Transformer} for speech
  recognition,''
\newblock in {\em Proc. of Interspeech}, 2020.

\bibitem{vaswani2017attention}
Ashish Vaswani et~al.,
\newblock ``Attention is all you need,''
\newblock in {\em Proc. of NeurIPS}, 2017, pp. 5998--6008.

\bibitem{gu2017non}
Jiatao Gu et~al.,
\newblock ``Non-autoregressive neural machine translation,''
\newblock in {\em Proc. of ICLR}, 2018.

\bibitem{libovicky2018end}
Jind{\v{r}}ich Libovick{\'y} and Jind{\v{r}}ich Helcl,
\newblock ``End-to-end non-autoregressive neural machine translation with
  connectionist temporal classification,''
\newblock in {\em Proc. of EMNLP}, 2018, pp. 3016--3021.

\bibitem{lee2018deterministic}
Jason Lee et~al.,
\newblock ``Deterministic non-autoregressive neural sequence modeling by
  iterative refinement,''
\newblock in {\em Proc. of EMNLP}, 2018, pp. 1173--1182.

\bibitem{stern2019insertion}
Mitchell Stern et~al.,
\newblock ``Insertion {Transformer}: Flexible sequence generation via insertion
  operations,''
\newblock in {\em Proc. of ICML}, 2019, pp. 5976--5985.

\bibitem{gu2019levenshtein}
Jiatao Gu et~al.,
\newblock ``Levenshtein {Transformer},''
\newblock in {\em Proc. of NeurIPS}, 2019, pp. 11181--11191.

\bibitem{ghazvininejad2019mask}
Marjan Ghazvininejad et~al.,
\newblock ``Mask-predict: Parallel decoding of conditional masked language
  models,''
\newblock in {\em Proc. of EMNLP-IJCNLP}, 2019, pp. 6114--6123.

\bibitem{ghazvininejad2020semi}
Marjan Ghazvininejad et~al.,
\newblock ``Semi-autoregressive training improves mask-predict decoding,''
\newblock {\em arXiv preprint arXiv:2001.08785}, 2020.

\bibitem{saharia2020non}
Chitwan Saharia et~al.,
\newblock ``Non-autoregressive machine translation with latent alignments,''
\newblock {\em arXiv preprint arXiv:2004.07437}, 2020.

\bibitem{ma2019flowseq}
Xuezhe Ma et~al.,
\newblock ``{FlowSeq}: Non-autoregressive conditional sequence generation with
  generative flow,''
\newblock in {\em Proc. of EMNLP-IJCNLP}, 2019, pp. 4273--4283.

\bibitem{graves2014towards}
Alex Graves and Navdeep Jaitly,
\newblock ``Towards end-to-end speech recognition with recurrent neural
  networks,''
\newblock in {\em Proceedings of ICML}, 2014, pp. 1764--1772.

\bibitem{chen2019non}
Nanxin Chen et~al.,
\newblock ``Listen and fill in the missing letters: Non-autoregressive
  {Transformer} for speech recognition,''
\newblock {\em arXiv preprint arXiv:1911.04908}, 2019.

\bibitem{chan2020imputer}
William Chan et~al.,
\newblock ``Imputer: Sequence modelling via imputation and dynamic
  programming,''
\newblock in {\em Proc. of ICML}, 2020.

\bibitem{higuchi2020mask}
Yosuke Higuchi et~al.,
\newblock ``Mask {CTC}: Non-autoregressive end-to-end {ASR} with {CTC} and mask
  predict,''
\newblock in {\em Proc. of Interspeech}, 2020.

\bibitem{fujita2020insertion}
Yuya Fujita et~al.,
\newblock ``Insertion-based modeling for end-to-end automatic speech
  recognition,''
\newblock in {\em Proc. of Interspeech}, 2020.

\bibitem{tian2020spike}
Zhengkun Tian et~al.,
\newblock ``Spike-triggered non-autoregressive {Transformer} for end-to-end
  speech recognition,''
\newblock in {\em Proc. of Interspeech}, 2020.

\bibitem{battenberg2017exploring}
Eric Battenberg et~al.,
\newblock ``Exploring neural transducers for end-to-end speech recognition,''
\newblock in {\em Proc. of ASRU}, 2017, pp. 206--213.

\bibitem{devlin2019bert}
Jacob Devlin et~al.,
\newblock ``{BERT}: Pre-training of deep bidirectional transformers for
  language understanding,''
\newblock in {\em Proc. of NAACL-HLT}, 2019, pp. 4171--4186.

\bibitem{kim2017joint}
Suyoun Kim et~al.,
\newblock ``Joint {CTC}-attention based end-to-end speech recognition using
  multi-task learning,''
\newblock in {\em Proc. of ICASSP}, 2017, pp. 4835--4839.

\bibitem{lu2019understanding}
Yiping Lu et~al.,
\newblock ``Understanding and improving transformer from a multi-particle
  dynamic system point of view,''
\newblock in {\em Proc. ICLR}, 2020.

\bibitem{ren2019fastspeech}
Yi~Ren et~al.,
\newblock ``{FastSpeech}: Fast, robust and controllable text to speech,''
\newblock in {\em Proc. of NeurIPS}, 2019.

\bibitem{watanabe2018espnet}
Shinji Watanabe et~al.,
\newblock ``{ESPnet}: End-to-end speech processing toolkit,''
\newblock in {\em Proc. of Interspeech}, 2018, pp. 2207--2211.

\bibitem{paul1992design}
Douglas~B Paul and Janet~M Baker,
\newblock ``The design for the wall street journal-based {CSR} corpus,''
\newblock in {\em Proc. of Workshop on Speech and Natural Language}, 1992, pp.
  357--362.

\bibitem{rousseau2014enhancing}
Anthony Rousseau et~al.,
\newblock ``Enhancing the {TED}-{LIUM} corpus with selected data for language
  modeling and more {TED} talks,''
\newblock in {\em Porc. of LREC}, May 2014, pp. 3935--3939.

\bibitem{voxforge}
``Voxforge,'' \url{http://www.voxforge.org}.

\bibitem{povey2011kaldi}
Daniel Povey et~al.,
\newblock ``The {Kaldi} speech recognition toolkit,''
\newblock in {\em Proc. of ASRU}, 2011.

\bibitem{ko2015audio}
Tom Ko et~al.,
\newblock ``Audio augmentation for speech recognition,''
\newblock in {\em Proc. of Interspeech}, 2015.

\bibitem{park2019specaugment}
Daniel~S Park et~al.,
\newblock ``{SpecAugment}: A simple data augmentation method for automatic
  speech recognition,''
\newblock in {\em Proc. of Interspeech}, 2019, pp. 2613--2617.

\bibitem{sennrich2016neural}
Rico Sennrich et~al.,
\newblock ``Neural machine translation of rare words with subword units,''
\newblock in {\em Proc. of ACL}, 2016, pp. 1715--1725.

\bibitem{hori2017joint}
Takaaki Hori et~al.,
\newblock ``Joint {CTC}/attention decoding for end-to-end speech recognition,''
\newblock in {\em Proc. of ACL}, 2017, pp. 518--529.

\bibitem{Sabour2019OptimalCD}
Sara Sabour et~al.,
\newblock ``Optimal completion distillation for sequence learning,''
\newblock in {\em Proc. of ICLR}, 2019.

\bibitem{borgholt2020do}
Lasse Borgholt et~al.,
\newblock ``Do end-to-end speech recognition models care about context?,''
\newblock in {\em Proc. of Interspeech}, 2020.

\bibitem{inaguma2020espnet}
Hirofumi Inaguma et~al.,
\newblock ``{ESPnet-ST}: All-in-one speech translation toolkit,''
\newblock in {\em Proc. ACL: System Demonstrations}, 2020, pp. 302--311.

\bibitem{kim2016sequence}
Yoon Kim et~al.,
\newblock ``Sequence-level knowledge distillation,''
\newblock in {\em Proc. of EMNLP}, 2016, pp. 1317--1327.

\bibitem{fisher_callhome}
Matt Post et~al.,
\newblock ``Improved speech-to-text translation with the {Fisher} and {Callhome
  Spanish--English} speech translation corpus,''
\newblock in {\em Proc. of IWSLT}, 2013.

\end{thebibliography}
\end{spacing}

\end{document}